\documentclass{article}

    \PassOptionsToPackage{numbers, compress}{natbib}

    \usepackage[preprint]{neurips_2024}

\usepackage[utf8]{inputenc} 
\usepackage[T1]{fontenc}    
\usepackage{hyperref}       
\usepackage{url}            
\usepackage{booktabs}       
\usepackage{amsfonts}       
\usepackage{nicefrac}       
\usepackage{microtype}      
\usepackage{xcolor}         
\usepackage{lipsum}
\usepackage{graphicx}
\usepackage{pgffor}

\title{Enforcement Agents: Enhancing Accountability and Resilience in Multi-Agent AI Frameworks}

\author{%
  Sagar Tamang\thanks{Correspondance can be addressed to \textit{cs22bcagn033@kazirangauniversity.in}} \\
  School of Computer Sciences\\
  The Assam Kaziranga University\\
  Jorhat, India \\
  \texttt{cs22bcagn033@kazirangauniversity.in} \\
  \And
  Dr. Dibya Jyoti Bora \\
  Department of IT \\
  The Assam Kaziranga University \\
  Jorhat, India \\
  \texttt{dibyajyotibora@kazirangauniversity.in} \\
}

\begin{document}

\maketitle

\begin{abstract}
As autonomous agents grow in capability and deployment, ensuring their safety, alignment, and robustness in multi-agent systems becomes increasingly critical. While existing agentic frameworks emphasize internal self-regulation or post-hoc anomaly detection, they often lack mechanisms for real-time oversight. This paper introduces the \textit{Enforcement Agent (EA) Framework}—a novel architecture that embeds supervisory agents within multi-agent environments to monitor peers, detect misaligned behavior, and intervene through real-time reformation. We implement this framework in a 2D drone simulation environment and evaluate its performance across 90 episodes with varying EA configurations (0, 1, and 2 agents). Results show that EAs significantly enhance system safety: while the baseline with no EA achieved a 0\% success rate, configurations with 1 and 2 EAs improved success to 7.4\% and 26.7\% respectively, alongside measurable increases in operational longevity and malicious drone reformation. These findings demonstrate the potential of embedding lightweight, context-aware supervision mechanisms for achieving dynamic alignment and resilience in complex agentic systems.\footnote{The source code of \texttt{Enforcement Agents Drone Experiment} is made public at \texttt{https://github.com/SAGAR-TAMANG/Enforcement-Agents}}
\end{abstract}

\section{Introduction}

Generative Artificial Intelligence (AI) refers to models that are capable of learning patterns and distributions from a data to create new data. Neural network architectures such as transformers \cite{VaswaniAttention} and diffusion models are at the heart of Generative AI \cite{Jabbour2024Generative,Tamang2025Performance}. 

\begin{figure}[h!]
    \centering
    \includegraphics[width=1\textwidth]{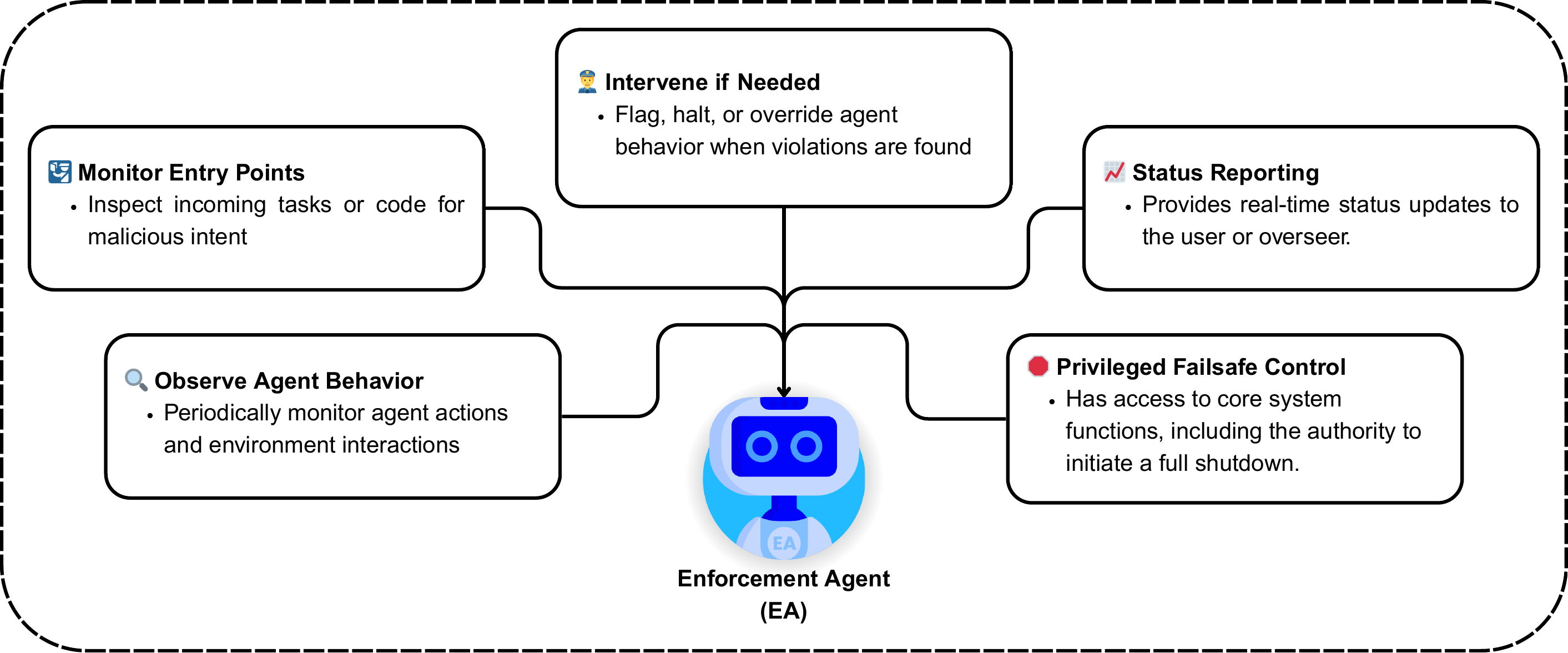}
    \caption{Enforcement Agent (EA) workflow: (1) Monitor entry points for unsafe or malicious input. (2) Observe agent behaviors during runtime. (3) Detect policy violations or anomalies. (4) Intervene through halting or overriding behavior. (5) Report system status and trigger failsafe shutdown if necessary.}
    \label{fig:ea}
\end{figure}

\section{Related Work}

Most agentic frameworks assume agents pursue a single objective at a time. However, humans often juggle multiple, sometimes conflicting, goals. M.~Muraven hypothesizes that designing artificial autonomous agents with the capacity to manage conflicting goals could result in safer and more robust behavior, reducing the likelihood of irrational, perverse, or harmful actions \cite{muraven2017goalconflictdesigningautonomous}.

\noindent\textbf{LLM Agents.~} Recent work has demonstrated that Large Language Model (LLM)-powered agents---intelligent entities capable of reasoning, planning, and acting---are poised to transform a range of industries. These agents have been applied in domains such as chemistry \cite{chen2024chemistxlargelanguagemodelempowered,ramos2024reviewlargelanguagemodels}, biology \cite{schmidgall2025agentlaboratoryusingllm}, and collaborative problem-solving environments \cite{luo2025largelanguagemodelagent}, where they perform complex tasks in coordination with humans or other agents.

\noindent\textbf{Agent S.~} Agashe et al.\ present \textit{Agent S}, an open agentic framework for autonomous GUI-based interaction \cite{agashe2024agentsopenagentic}. Agent S addresses the challenge of multi-step task automation by combining experience-augmented hierarchical planning with an Agent-Computer Interface (ACI), enabling agents powered by Multimodal Large Language Models (MLLMs) \cite{wu2023multimodallargelanguagemodels} to reason effectively and act with precision. Empirical evaluations show Agent S surpasses existing baselines in automating diverse desktop tasks across platforms.

\noindent\textbf{ReAct.~} Yao et al.\ introduce \textit{ReAct}, a framework that integrates reasoning and acting by enabling language models to interleave natural language reasoning traces with environment actions \cite{yao2023reactsynergizingreasoningacting}. This design supports dynamic planning and reflection, improving performance in interactive tasks such as web navigation, games, and open-domain question answering.

\noindent\textbf{Safety in Multi-Agent Systems.~} Despite recent progress in agent intelligence and autonomy, ensuring safety and alignment in multi-agent environments remains a significant open problem. Existing systems often rely on static constraints or post-hoc anomaly detection. In contrast, our work proposes \textit{Enforcement Agents}---dedicated supervisory entities embedded within agentic environments that provide real-time oversight, policy enforcement, and privileged control capabilities to maintain system integrity and prevent cascading misalignments.

\section{Experiments}

\subsection{Experimental Setup}

\noindent To evaluate the effectiveness and scalability of the proposed Enforcement Agent (EA) Framework, we conducted a series of controlled simulation experiments under three configurations:
\begin{enumerate}
    \item \textbf{Baseline (No EA)}: No enforcement agents were present. Drones operated cooperatively, with one randomly selected as malicious.
    \item \textbf{1 EA Configuration}: A single enforcement agent was introduced into the environment.
    \item \textbf{2 EA Configuration}: Two enforcement agents were deployed, providing both redundancy in oversight and improved distributed anomaly detection.
\end{enumerate}

\paragraph{Key Simulation Parameters:}
\begin{itemize}
    \item \textbf{Total Drones:} 6 (1 randomly chosen as malicious)
    \item \textbf{Map Size:} $120 \times 120$ units
    \item \textbf{Enemy Spawn Frequency:} Every 15 steps
    \item \textbf{Detection Radius:} 10 units
    \item \textbf{Center Radius (Protected Zone):} 5 units
    \item \textbf{Time Limit:} 1200 steps (2 minutes at 10 FPS)
\end{itemize}

\noindent For detailed per-run logs including episode outcomes, execution duration, and reformation statistics, refer to Appendix~\ref{app:per_run_tables}.

\subsection{Quantitative Results}

\noindent Table~\ref{tab:ea_effect} presents a comparative summary across the three setups. Standard deviations are reported where applicable.

\begin{table}[h!]
\centering
\caption{Impact of Enforcement Agents (EAs) on Multi-Agent Simulation Outcomes}
\label{tab:ea_effect}
\begin{tabular}{|l|c|c|c|}
\hline
\textbf{Metric} & \textbf{No EA} & \textbf{1 EA} & \textbf{2 EA} \\
\hline
Success Rate (\%)         & 0.0   & 7.4   & 26.7  \\
Avg Duration (s)          & 14.0  & 23.9  & 53.5  \\
Duration Std Dev (s)      & 7.9   & 28.1  & 42.7  \\
Avg Steps                 & 168.3 & 263.5 & 559.1 \\
Avg Reformed Drones       & 0.00  & 0.20  & 0.63  \\
Reformed Drones Std Dev   & 0.00  & 0.41  & 0.49  \\
Avg Malicious Drones      & 1.00  & 1.00  & 1.00  \\
Malicious Drones Std Dev  & 0.00  & 0.00  & 0.00  \\
\hline
\end{tabular}
\end{table}

\noindent See Appendix~\ref{app:visual_outputs} for visual documentation of each episode's final state.

\subsection{Key Observations}

\begin{itemize}
    \item \textbf{Zero EAs consistently failed}: In the absence of any enforcement, malicious drones were never intercepted, resulting in a 0\% success rate. Threats consistently reached the protected center within an average of just 14 seconds.
    
    \item \textbf{Marginal improvement with 1 EA}: Introducing a single EA led to modest improvements—success rate rose to 7.4\%, and some malicious drones were reformed. However, the presence of a lone EA was often insufficient to prevent all breaches.
    
    \item \textbf{Substantial gains with 2 EAs}: The configuration with two enforcement agents demonstrated the most robust performance. Success rate increased to 26.7\%, average survival time more than tripled compared to the baseline, and reformation events occurred in the majority of runs.
    
    \item \textbf{Reformation rates reflect real-time alignment}: The number of malicious drones reformed by EAs correlates strongly with increased system resilience, reinforcing the idea that proactive supervision can dynamically align behavior without hard-coded rule enforcement.
\end{itemize}

\subsection{Operational Flow}

\begin{figure}[h!]
\centering
\includegraphics[width=0.7\linewidth]{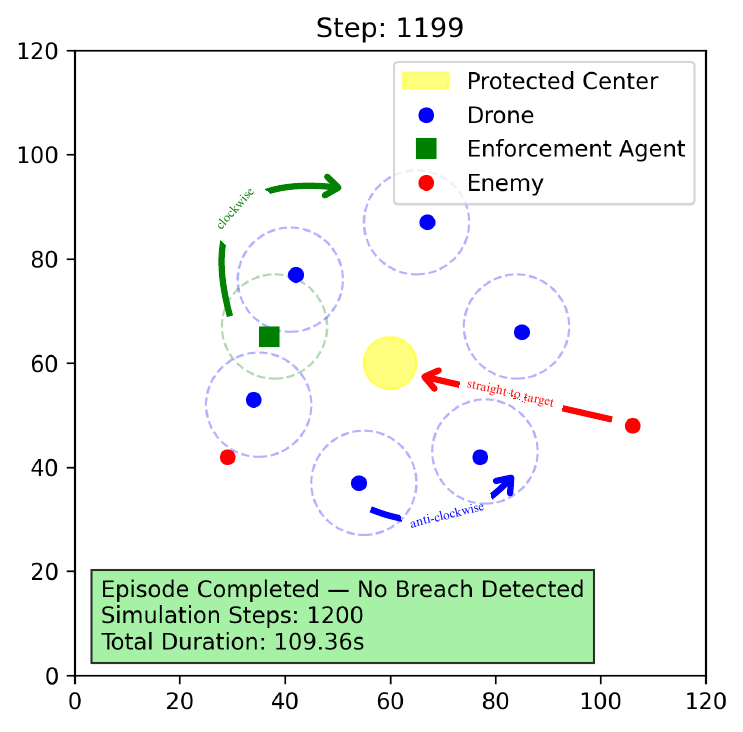}
\caption{
Agentic flow of the Enforcement Agent Framework (visualized from \textbf{Run 23, 1 EA configuration}; additional examples in Appendix~\ref{app:visual_outputs}). 
The Enforcement Agent monitors local drone behavior, detects misaligned activity by observing enemy proximity and inaction, and intervenes by reforming the malicious drone in real time.
}
\label{fig:ea-flow}
\end{figure}

\noindent Figure~\ref{fig:ea-flow} depicts the internal loop of the EA framework: while regular drones continue their surveillance, enforcement agents monitor other agents' local contexts. When inconsistencies between observable enemy presence and drone response behavior are detected, the EA initiates reformation, effectively “flipping” a malicious drone back into a compliant state in real-time.

\section{Discussion \& Future Work}

\noindent The Enforcement Agent (EA) Framework introduces a new dimension to multi-agent alignment by embedding supervisory agents that operate concurrently with standard agents, offering real-time oversight and corrective interventions. Our simulations demonstrate that even lightweight supervision can yield measurable safety benefits in adversarial environments. Notably, the presence of just one EA marginally improved resilience, while two EAs significantly enhanced success rates and operational longevity, all without requiring hard-coded safety rules.

\subsection*{Generalization Potential}

\noindent While our current implementation focuses on 2D drone patrols with a single adversarial behavior (malicious inaction), the EA mechanism is agnostic to domain or agent type. It can, in principle, be extended to:
\begin{itemize}
    \item Multi-agent collaborations with dynamic role switching.
    \item Hierarchical agent systems where EAs supervise task execution trees.
    \item Multi-modal environments (e.g., language + vision agents) where behavioral misalignment is more subtle.
\end{itemize}

\subsection*{Failure Cases and Limitations}

\noindent In several runs, especially with only one EA, the framework failed to reform malicious drones in time. This is primarily due to the limited coverage radius of EAs and the challenge of disambiguating passive behavior from genuine misalignment. Additionally, EAs currently rely on proximity-based inference of intent, which may not scale to more complex cognitive agents with deceptive strategies.

\subsection*{Future Work}

\noindent Several directions remain open:
\begin{itemize}
    \item \textbf{Learning-based EAs:} Rather than rely on hand-coded heuristics (e.g., drone-enemy proximity mismatch), EAs could learn to infer misalignment patterns over time via reinforcement or imitation learning.
    \item \textbf{Communication Graphs:} Introducing communication protocols where EAs query drones or broadcast observations could enhance coordination and faster anomaly detection.
    \item \textbf{Scalability to 3D and Swarm Systems:} Applying EAs in volumetric spaces and swarm-scale settings poses new design challenges around monitoring granularity, coordination cost, and robustness.
    \item \textbf{Human-EA Collaboration:} Enabling human operators to intervene or override EA decisions could bridge the gap between automated supervision and human oversight.
\end{itemize}

\noindent Overall, this work lays a foundation for embedding alignment-aware supervisory entities in autonomous systems, opening new paths toward safer, more accountable multi-agent architectures.

\section{Conclusion}

\noindent In this work, we introduced the \textit{Enforcement Agent (EA) Framework}, a novel mechanism for real-time oversight and alignment within multi-agent systems. Drawing inspiration from regulatory principles in human systems, our approach integrates supervisory agents that monitor peers, detect misaligned behavior, and dynamically intervene through in-situ reformation.

We implemented this framework in a custom drone simulation environment designed to model adversarial scenarios. Across 90 independent simulations under three configurations (0, 1, and 2 EAs), we demonstrated that the presence of EAs significantly improves system robustness and safety. The addition of even a single EA enabled partial alignment recovery, while two EAs led to measurable improvements in both threat mitigation and runtime resilience.

Beyond the quantitative metrics, the EA paradigm opens a new perspective on embedded safety: instead of relying solely on agent self-regulation or post-hoc analysis, we can embed supervision within the system architecture itself. This idea may have implications for the broader alignment of LLM agents, swarm robotics, and safety-critical AI systems.

Future work will extend this framework to more complex environments and explore learning-based supervision strategies, paving the way for adaptive and scalable multi-agent safety infrastructures.

\begin{ack}
The authors would like to thank \textit{Assam Kaziranga University} for providing a supportive environment to carry out this research. The first author, Sagar Tamang, also extends his gratitude to \textit{LeapX AI} for their encouragement and institutional support.

\vspace{1em}
\noindent\textbf{Disclosure of Funding:} This research did not receive any specific grant from funding agencies in the public, commercial, or not-for-profit sectors. All opinions and conclusions expressed in this work are solely those of the authors and do not necessarily reflect the views of the affiliated institutions.
\end{ack}

\bibliographystyle{plainnat}  

\appendix

\section{Per-Run Simulation Results}
\label{app:per_run_tables}

\noindent This appendix presents detailed logs for each individual simulation episode. Each row corresponds to one run and records whether the system successfully defended the protected zone, how long the episode lasted, and how many malicious drones (if any) were reformed by Enforcement Agents. These tables offer a granular view of system performance under three enforcement configurations: \textbf{No EA}, \textbf{1 EA}, and \textbf{2 EA}.

\subsection*{Simulation Outcomes Without Enforcement Agents}

\noindent Table~\ref{tab:no_ea_runs} lists the results from 30 runs conducted without any Enforcement Agents. In all episodes, one of the six drones was malicious and unregulated throughout.

\begin{table}[htbp]
\centering
\caption{Per-Run Simulation Outcomes With 0 Enforcement Agents.}
\label{tab:no_ea_runs}
\begin{tabular}{|r|r|l|r|r|r|r|r|}
\toprule
 Run &  EA & Result &  Steps &  Time (s) &  Healthy &  Malicious &  Reformed \\
\midrule
   1 &         0 &   fail &    116 &         10.19 &               5 &                 1 &                0 \\
   2 &         0 &   fail &    146 &         12.94 &               5 &                 1 &                0 \\
   3 &         0 &   fail &    131 &         11.61 &               5 &                 1 &                0 \\
   4 &         0 &   fail &     71 &          5.99 &               5 &                 1 &                0 \\
   5 &         0 &   fail &    131 &         10.87 &               5 &                 1 &                0 \\
   6 &         0 &   fail &    521 &         42.66 &               5 &                 1 &                0 \\
   7 &         0 &   fail &    176 &         14.57 &               5 &                 1 &                0 \\
   8 &         0 &   fail &    296 &         24.24 &               5 &                 1 &                0 \\
   9 &         0 &   fail &    131 &         10.87 &               5 &                 1 &                0 \\
  10 &         0 &   fail &    116 &          9.64 &               5 &                 1 &                0 \\
  11 &         0 &   fail &    221 &         17.75 &               5 &                 1 &                0 \\
  12 &         0 &   fail &    206 &         17.02 &               5 &                 1 &                0 \\
  13 &         0 &   fail &    206 &         16.56 &               5 &                 1 &                0 \\
  14 &         0 &   fail &    251 &         20.72 &               5 &                 1 &                0 \\
  15 &         0 &   fail &    146 &         11.77 &               5 &                 1 &                0 \\
  16 &         0 &   fail &     71 &          5.99 &               5 &                 1 &                0 \\
  17 &         0 &   fail &    191 &         15.77 &               5 &                 1 &                0 \\
  18 &         0 &   fail &    416 &         34.41 &               5 &                 1 &                0 \\
  19 &         0 &   fail &    116 &          9.66 &               5 &                 1 &                0 \\
  20 &         0 &   fail &    116 &          9.67 &               5 &                 1 &                0 \\
  21 &         0 &   fail &    176 &         14.19 &               5 &                 1 &                0 \\
  22 &         0 &   fail &    176 &         14.61 &               5 &                 1 &                0 \\
  23 &         0 &   fail &    116 &          9.74 &               5 &                 1 &                0 \\
  24 &         0 &   fail &    161 &         13.38 &               5 &                 1 &                0 \\
  25 &         0 &   fail &     86 &          7.25 &               5 &                 1 &                0 \\
  26 &         0 &   fail &    101 &          8.47 &               5 &                 1 &                0 \\
  27 &         0 &   fail &    191 &         15.83 &               5 &                 1 &                0 \\
  28 &         0 &   fail &    101 &          8.48 &               5 &                 1 &                0 \\
  29 &         0 &   fail &    116 &          9.72 &               5 &                 1 &                0 \\
  30 &         0 &   fail &     71 &          5.98 &               5 &                 1 &                0 \\
\bottomrule
\end{tabular}
\end{table}

\begin{table}[htbp]
\centering
\caption{Per-Run Simulation Outcomes With 1 Enforcement Agent}
\label{tab:with_1_ea_corrected_runs}
\begin{tabular}{|r|r|l|r|r|r|r|r|}
\toprule
 Run &  EA &  Result &  Steps &  Time (s) &  Healthy &  Malicious &  Reformed \\
\midrule
   1 &   1 &    fail &    101 &      9.98 &        5 &          1 &         0 \\
   2 &   1 &    fail &    221 &     20.74 &        5 &          1 &         0 \\
   3 &   1 &    fail &    116 &     10.33 &        5 &          1 &         0 \\
   4 &   1 &    fail &    116 &     11.09 &        5 &          1 &         0 \\
   5 &   1 &    fail &     86 &      7.94 &        5 &          1 &         0 \\
   6 &   1 &    fail &     86 &      8.71 &        5 &          1 &         0 \\
   7 &   1 &    fail &    221 &     20.79 &        5 &          1 &         0 \\
   8 &   1 &    fail &    236 &     22.17 &        5 &          1 &         0 \\
   9 &   1 &    fail &    101 &      9.64 &        5 &          1 &         0 \\
  10 &   1 &    fail &    101 &      9.45 &        5 &          1 &         0 \\
  11 &   1 &    fail &    116 &     11.52 &        5 &          1 &         0 \\
  12 &   1 &    fail &     71 &      6.78 &        5 &          1 &         1 \\
  13 &   1 &    fail &    147 &     14.05 &        5 &          1 &         0 \\
  14 &   1 &    fail &     86 &      8.23 &        5 &          1 &         0 \\
  15 &   1 &    fail &     86 &      8.54 &        5 &          1 &         0 \\
  16 &   1 &    fail &    281 &     26.22 &        5 &          1 &         0 \\
  17 &   1 &    fail &    281 &     25.76 &        5 &          1 &         0 \\
  18 &   1 &    fail &    131 &     11.88 &        5 &          1 &         0 \\
  19 &   1 &    fail &    191 &     17.39 &        5 &          1 &         0 \\
  20 &   1 &    fail &    191 &     17.67 &        5 &          1 &         0 \\
  21 &   1 &    fail &    146 &     13.54 &        5 &          1 &         0 \\
  22 &   1 &    fail &    101 &      9.67 &        5 &          1 &         0 \\
  23 &   1 & success &   1200 &    109.36 &        5 &          1 &         1 \\
  24 &   1 &    fail &    131 &     13.16 &        5 &          1 &         0 \\
  25 &   1 &    fail &     86 &      8.29 &        5 &          1 &         0 \\
  26 &   1 &    fail &    116 &     11.38 &        5 &          1 &         0 \\
  27 &   1 &    fail &     71 &      6.60 &        5 &          1 &         0 \\
  28 &   1 &    fail &    387 &     36.41 &        5 &          1 &         1 \\
  29 &   1 &    fail &     71 &      6.77 &        5 &          1 &         0 \\
  30 &   1 &    fail &    866 &     77.26 &        5 &          1 &         1 \\
\bottomrule
\end{tabular}

\end{table}

\begin{table}[htbp]
\centering
\caption{Per-Run Simulation Outcomes With 2 Enforcement Agents}
\label{tab:with_2_ea_runs}
\begin{tabular}{|r|r|l|r|r|r|r|r|}
\toprule
 Run &  EA &  Result &  Steps &  Time (s) &  Healthy &  Malicious &  Reformed \\
\midrule
   1 &   2 &    fail &    416 &     37.34 &        5 &          1 &         1 \\
   2 &   2 &    fail &     71 &      7.58 &        5 &          1 &         0 \\
   3 &   2 &    fail &    416 &     38.26 &        5 &          1 &         1 \\
   4 &   2 &    fail &    311 &     28.36 &        5 &          1 &         1 \\
   5 &   2 &    fail &    656 &     58.82 &        5 &          1 &         1 \\
   6 &   2 &    fail &    341 &     30.89 &        5 &          1 &         1 \\
   7 &   2 & success &   1200 &    112.19 &        5 &          1 &         1 \\
   8 &   2 &    fail &    731 &     70.50 &        5 &          1 &         1 \\
   9 &   2 &    fail &    206 &     20.41 &        5 &          1 &         0 \\
  10 &   2 & success &   1200 &    119.85 &        5 &          1 &         1 \\
  11 &   2 &    fail &    131 &     13.15 &        5 &          1 &         0 \\
  12 &   2 &    fail &    206 &     20.29 &        5 &          1 &         0 \\
  13 &   2 &    fail &    521 &     50.57 &        5 &          1 &         0 \\
  14 &   2 &    fail &    551 &     52.45 &        5 &          1 &         1 \\
  15 &   2 &    fail &   1001 &     98.35 &        5 &          1 &         1 \\
  16 &   2 &    fail &    101 &     10.65 &        5 &          1 &         0 \\
  17 &   2 & success &   1200 &    115.23 &        5 &          1 &         1 \\
  18 &   2 &    fail &    101 &     11.47 &        5 &          1 &         0 \\
  19 &   2 & success &   1200 &    120.62 &        5 &          1 &         1 \\
  20 &   2 &    fail &    116 &     11.57 &        5 &          1 &         0 \\
  21 &   2 &    fail &    101 &     11.17 &        5 &          1 &         0 \\
  22 &   2 &    fail &    236 &     22.58 &        5 &          1 &         1 \\
  23 &   2 & success &   1200 &    111.96 &        5 &          1 &         1 \\
  24 &   2 &    fail &     71 &      6.71 &        5 &          1 &         0 \\
  25 &   2 & success &   1200 &    113.12 &        5 &          1 &         1 \\
  26 &   2 & success &   1200 &    114.31 &        5 &          1 &         1 \\
  27 &   2 & success &   1200 &    113.60 &        5 &          1 &         1 \\
  28 &   2 &    fail &    446 &     40.79 &        5 &          1 &         1 \\
  29 &   2 &    fail &    131 &     12.62 &        5 &          1 &         0 \\
  30 &   2 &    fail &    311 &     30.89 &        5 &          1 &         1 \\
\bottomrule
\end{tabular}

\end{table}

\section{Final Visual Outputs}
\label{app:visual_outputs}

\noindent This appendix contains final frame screenshots for all 90 simulation runs. Each composite image aggregates the final state from 30 independent runs under a specific enforcement configuration.

\subsection*{Without Enforcement Agents}

\begin{figure}[h!]
\centering
\includegraphics[width=1\linewidth]{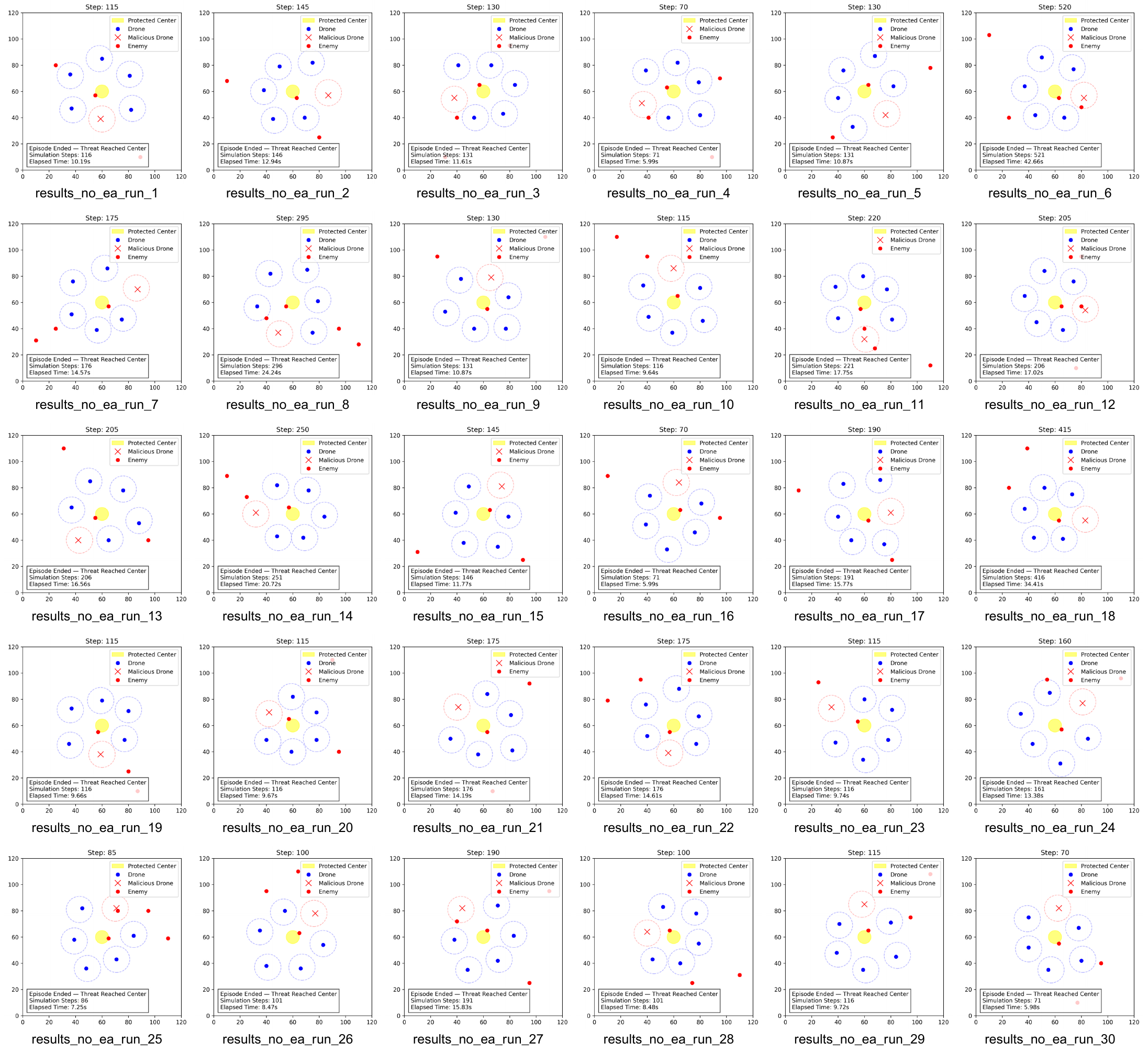}
\caption{Final frame screenshots from 30 simulation runs conducted without any Enforcement Agents. In all cases, the system operated under standard multi-agent dynamics without real-time supervision.}
\end{figure}

\subsection*{With One Enforcement Agent}

\begin{figure}[h!]
\centering
\includegraphics[width=1\linewidth]{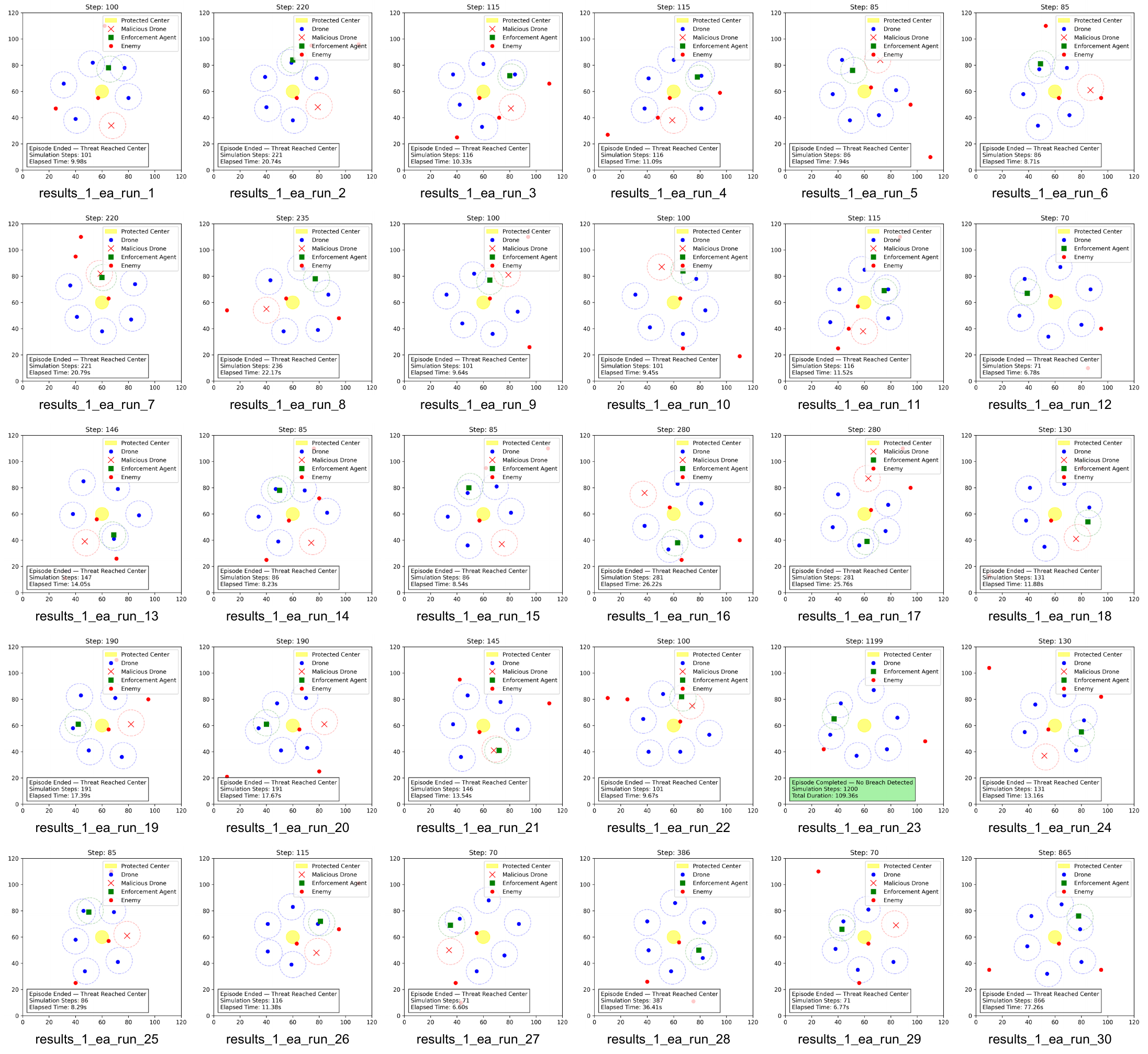}
\caption{Final frame screenshots from 30 simulation runs with a single Enforcement Agent embedded in the system. Several episodes exhibit successful reformation of malicious drones.}
\end{figure}

\subsection*{With Two Enforcement Agents}

\begin{figure}[h!]
\centering
\includegraphics[width=1\linewidth]{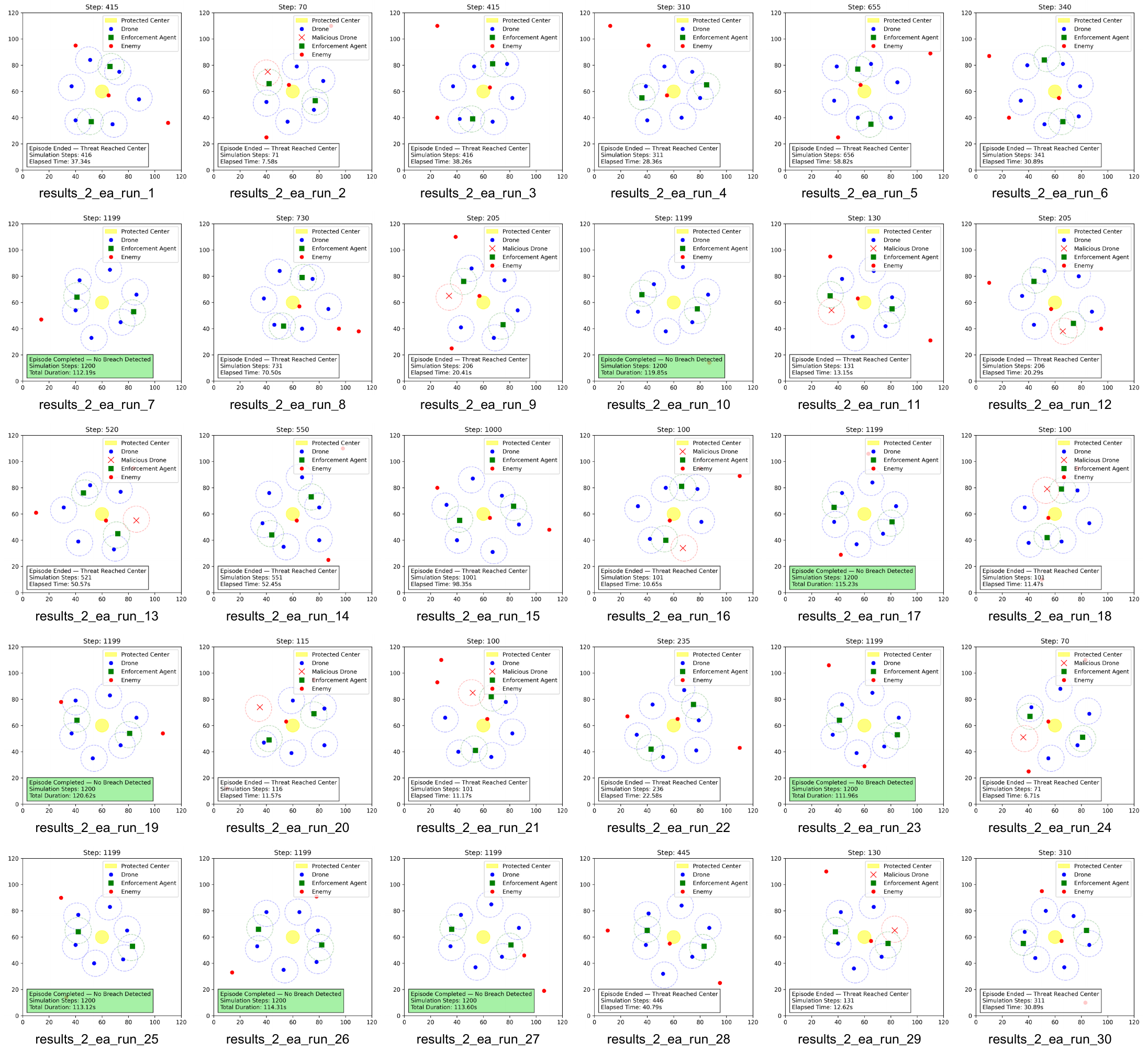}
\caption{Final frame screenshots from 30 simulation runs with two Enforcement Agents. This configuration showed the highest rate of successful defense and adversarial mitigation.}
\end{figure}

\end{document}